\documentstyle[preprint,prc,eqsecnum,aps]{revtex}
\begin{document}
\draft
\def\lsim{\thinspace{\hbox to 8pt{\raise
-5pt\hbox{$\sim$}\hss{$<$}}}\thinspace}
\def\rsim{\thinspace{\hbox to 8pt{\raise
-5pt\hbox{$\sim$}\hss{$>$}}}\thinspace}
\title{
Consistent Treatment of Propagator Modifications in \\
Elastic Nucleon-Nucleus
Scattering within the Spectator Expansion}

\author{ C.R.~Chinn$^{(a),(b)}$, Ch.~Elster$^{(c)}$,
 R.M.~Thaler$^{(a),(d)}$, and S.P.~Weppner$^{(c)}$.}
\address{
$^{(a)}$ Department of Physics and Astronomy, Vanderbilt University,
Nashville, TN  37235}
\address{
$^{(b)}$ Center for Computationally Intensive Physics,
Oak Ridge National Laboratory, \\ Oak Ridge, TN  37831}
\address{
$^{(c)}$ Institute of Nuclear and Particle Physics,  and
Department of Physics, \\ Ohio University, Athens, OH 45701}
\address{
$^{(d)}$ Physics Department, Case Western Reserve University, Cleveland,
OH  44106.}

\vspace{10mm}

\date{\today}

\maketitle

\begin{abstract}

The theory of the elastic scattering of a nucleon from a nucleus
is presented in the form of a Spectator Expansion of the optical
potential.  Particular attention is paid to the treatment of the
free projectile$\,-\,$nucleus propagator when the coupling of
the struck target nucleon to the residual target must be taken
into consideration.  First order calculations within this
framework are shown for neutron total cross-sections and for
proton scattering for a number of target nuclides at a variety of
energies.  The calculated values of these observables are in very good
agreement with measurement.

\end{abstract}

\vspace{10mm}


\pagebreak

%****************************************************************************

\narrowtext

%******************************************************************

\section{Introduction}

\hspace*{10mm}
The theoretical approach to the elastic scattering of a nucleon from a
nucleus, pioneered by Watson \cite{Watson}, made familiar by
Kerman, McManus, and Thaler (KMT) \cite{KMT} and further developed as
the Spectator Expansion \cite{Corr,Sicil,TT} is now being applied with
striking success.  In a similar vain, a slightly different
approach to the multiple scattering expansion within the KMT framework
is being pursued by the Surrey group \cite{Crespo}.

\hspace*{10mm}
The theoretical motivation for the Spectator
Expansion derives from our present inability to calculate the full
many-body problem.  In this case an expansion is constructed
within a consistent multiple scattering theory predicated upon the
idea that two-body interactions between the projectile and the
target nucleons inside the nucleus play the dominant role.
In the Spectator Expansion the first order term involves
two-body interactions between the projectile and one of the target
nucleons, the second order term involves the projectile interacting
with two target nucleons and so forth.  Hence the expansion
derives the ordering from the number of target nucleons
interacting directly with the projectile, while the residual target
nucleus remains `passive'.
Due to the many-body nature of the free propagator for the
the projectile$\,+\,$target system it is necessary to detail certain
choices made with respect to the ordering in the Spectator series.
Presented in this paper are the details of the Spectator Expansion
and the present manner in which the first order theory is
calculated, including a theoretical treatment of the many-body
propagator as affected by the residual target nucleus.
Predictions are shown for rigorous calculations
of the elastic scattering of protons and neutrons
from a variety of target nuclei in the energy regime between 65 and
400 MeV.  The calculated observables are in very good
agreement with the experimental information within this energy regime,
indicating the success that this theory enjoys.  It is very satisfying
to observe in this completely consistent theoretical framework,
that as the sophistication of the calculation is increased, the
resulting predictions invariably improve.

\hspace*{10mm}
The calculation of the multiple scattering theory as presented in
this paper relies on two basic inputs.  One is the fully off-shell
nucleon-nucleon (NN) t-matrix, which represents the best current
understanding of the nuclear force, and the other is
the nuclear wave function of the target, representing the best present
understanding of the ground state of the target nucleus.
These quantities comprise the required physical ingredients for
a microscopic construction of an optical potential for elastic
scattering.  To account for the modifications of the free
propagator inside the nucleus, mean field potentials taken from
microscopic nuclear structure calculations are used.
It must be emphasized that there are {\underline {no}}
adjustable parameters present in these calculations.

\hspace*{10mm}
The motivation for ongoing work on this topic is twofold. First,
elastic and inelastic nucleon-nucleus scattering provide an
important and sensitive test for theoretical corrections at the
first-order level of the optical potential ({\it e.g.} as
given by possibly genuine modifications
of the NN interaction in the nuclear environment
and off-shell effects).  Rigorous microscopic
calculations are required for discerning these effects.  A clear
understanding of this theory is also necessary before steps can be
taken to address the next level of sophistication.
Second, a better understanding of the theoretical details of the
optical potential are needed to construct realistic and physically
sound wave functions representing continuum nucleons in the
interior of the nucleus.  These wave functions will become vital
for future theoretical needs in high-energy coincidence experiments
at CEBAF, inelastic scattering studies,
and for understanding the reactions in heavy ion experiments
involving the new generation of radioactive beam facilities.
(In one sense to be able to develop a microscopic scattering theory
for heavy ions it is necessary to first clarify the
multiple scattering theory of hadronic probes.)

\hspace*{10mm}
The theoretical framework is presented in Section~II, namely
the Spectator Expansion, the first order term, the modification of
the propagator due to the residual spectator nucleons and the
second order term.  Section~III provides the details of the
calculations and the results for neutron-nucleus and
proton-nucleus elastic scattering.  A conclusion follows in
Section~IV.

\section{Theoretical Framework}
\subsection{The Spectator Expansion}

\hspace*{10mm}
The basic motivation behind the Spectator Expansion is that the
solution of the full
many-body problem is beyond present capabilities, hence an
expansion series is constructed for multiple scattering theory
predicated upon the number of target nucleons interacting directly
with the projectile.   Hence the expansion involves terms
where the projectile interacts directly with one target nucleon
plus a second order term where the projectile interacts directly
with two target nucleons, and so on to third and subsequent orders.
The separation of these terms with
respect to these categories of interactions is not completely
fixed due to the nature of the complicated $A+1$ body
propagator, hence some possible choices detailed in this paper must be
differentiated.

\hspace*{10mm}
At the heart of the standard approach to the elastic scattering of a
single projectile from a target of $A$ particles is the separation
into two parts of the Lippmann-Schwinger equation for the the
transition operator $T$, as given by
\begin{equation}
T = V + V G_0(E) T.  \label{eq:2.1}
\end{equation}
These two parts are an integral equation for $T$,
\begin{equation}
T = U + U G_0(E) P T,  \label{eq:2.2}
\end{equation}
where here $U$ is the optical potential operator, and an integral
equation for $U$
\begin{equation}
U = V + V G_0(E) Q U.  \label{eq:2.3}
\end{equation}
In the above equations the operator V represents the
external interaction, such that
the Hamiltonian for the entire $A+1$ particle system is given by
\begin{equation}
H=H_0 + V .   \label{eq:2.4}
\end{equation}
Asymptotically the system is in an eigenstate of $H_0$, and the free
propagator $G_0(E)$ for the projectile$\,+\,$target nucleus system is
\begin{equation}
G_0(E) = (E - H_0 + i\varepsilon)^{-1} . \label{eq:2.5}
\end{equation}
The operators $P$ and $Q$ are projection operators, $P+Q=1$
and $P$ is defined such that Eq.~(\ref{eq:2.2}) is solvable.  In this case
$P$ is conveniently taken to project onto the elastic channel, such
that, among other properties we have
\begin{equation}
[G_0,P] = 0 . \label{eq:2.6}
\end{equation}

For the scattering of a single particle projectile from an $A$-particle
target the free Hamiltonian is given by
\begin{equation}
H_0 = h_0 +H_A , \label{eq:2.7}
\end{equation}
where $h_0$ is the kinetic energy operator for the projectile and
$H_A$ stands for the target Hamiltonian.
Thus the projector $P$ can be defined as
\begin{equation}
P= \frac{|\Phi_A \rangle \langle \Phi_A|}
 {\langle \Phi_A|\Phi_A \rangle}, \label{eq:2.8}
\end{equation}
where $|\Phi_A \rangle$ corresponds to the ground state of the target,
satisfying the condition given in Eq.~(\ref{eq:2.6}), and
fulfilling
\begin{equation}
H_A |\Phi_A \rangle = E_A |\Phi_A \rangle . \label{eq:2.9}
\end{equation}

With these definitions the transition operator for elastic scattering
may be defined as $T_{el} = PTP$, in which case
Eq.~(\ref{eq:2.2})  can be written as
\begin{equation}
T_{el} = PUP + PUP G_0(E) T_{el}.  \label{eq:2.12}
\end{equation}
Thus, the transition operator for elastic scattering is given by
a straightforward one-body integral equation, which requires, of course,
the knowledge of the operator $PUP$.  The theoretical treatment which
follows consists of a formulation of the many-body
equation, Eq. (\ref{eq:2.3}),  where expressions for $U$ are
derived such that $PUP$ can be calculated accurately without
having to solve the complete many-body problem.

\hspace*{10mm}
For the present discussion, the presence of two-body forces
only is assumed. The extension to $A$-body forces is
straightforward. With this assumption the operator $U$ for the
optical potential can be expressed as
\begin{equation}
U = \sum_{i=1}^A U_i  \label{eq:2.13}
\end{equation}
where $U_i$ is given by
\begin{equation}
U_i = v_{0i} + v_{0i} G_0(E) Q \sum_{j=1}^A U_j ,  \label{eq:2.14}
\end{equation}
provided that
\begin{equation}
V= \sum_{i=1}^A v_{0i}.
\end{equation}
The two-body potential, $v_{0i}$, acts between the
projectile and the $i$th target nucleon.
Through the introduction of an operator $\tau_i$ which satisfies
\begin{equation}
\tau_i = v_{0i} + v_{0i} G_0(E) Q \tau_i , \label{eq:2.15}
\end{equation}
Eq.~(\ref{eq:2.14}) can be rearranged as
\begin{equation}
U_i = \tau_i + \tau_i G_0(E) Q \sum_{j \neq i} U_j . \label{eq:2.16}
\end{equation}
This rearrangement process can be continued for all $A$ target particles,
so that the operator for the optical potential can be expanded in  a series
of $A$ terms of the form
\begin{equation}
U = \sum_{i=1}^A \tau_i + \sum_{i,j \neq i}^A \tau_{ij} +
    \sum_{i,j \neq i,k \neq i,j}^A \tau_{ijk} + \cdots . \label{eq:2.17}
\end{equation}
This is the Spectator Expansion, where each term is treated in turn.
The separation of the interactions according to the number of
interacting nucleons has a certain latitude, due to the
many-body nature of $G_0(E)$.

We now concentrate on $\tau_{ij}$, which appears in the second term of
Eq.~(\ref{eq:2.17}).
Its ingredients are readily obtained from Eq.~(\ref{eq:2.14})
by means of the definition
\begin{equation}
(U_i - \tau_i) \equiv \sum_{j \neq i} \xi_{ij}. \label{eq:2.18}
\end{equation}
The operator $\xi_{ij}$, so defined, satisfies the
following many-body integral equation
\begin{eqnarray}
\xi_{ij} &=& \tau_i G_0(E) Q \tau_j + \sum_{k \neq j} \tau_i G_0(E) Q \xi_{jk}
                                               \nonumber \\
      &=& \tau_i G_0(E) Q \tau_j + \tau_i G_0(E) Q \xi_{ji} +
                                  \tau_i G_0(E) Q \sum_{k \neq i,j} \xi_{jk}
                                               \nonumber \\
    &=& \tau_i G_0(E) Q \tau_j + \tau_i G_0(E) Q \xi_{ji} + O (i,j,k).
                                                \label{eq:2.19}
\end{eqnarray}
Omitting all (i,j,k) terms on the right hand side of Eq.~(\ref{eq:2.19})
leads to the second term in Eq.~(\ref{eq:2.17}) via the
identification of $\tau_{ij}$ as
\begin{equation}
\tau_{ij} = \tau_i G_0(E) Q \tau_j + \tau_i G_0(E) Q \tau_{ji}. \label{eq:2.20}
\end{equation}

\hspace*{10mm}
The physical interpretation of Eq.~(\ref{eq:2.20}) can be most easily
recognized through an operator $\chi_{ij}$, defined as
\begin{equation}
\chi_{ij} = \tau_i + \tau_{ij} , \label{eq:2.21}
\end{equation}
from which the following relation is obtained:
\begin{equation}
\chi_{ij} = \tau_i +  \tau_i G_0(E) Q \chi_{ji}. \label{eq:2.22}
\end{equation}
{}From the symmetric combination,
$\tilde{U}_{ij}\equiv\chi_{ij}  + \chi_{ji}$, a standard
three-body equation is derived:
\begin{equation}
\tilde{U}_{ij}=(v_{0i}+v_{0j})+(v_{0i}+v_{0j}) G_0(E) Q \tilde{U}_{ij}.
                                 \label{eq:2.23}
\end{equation}

\hspace*{10mm}
The finite series given in Eq~(\ref{eq:2.17}) together with the
definitions of $\tau_i$, $\tau_{ij}$, $\cdots$ given above constitute one
form of the Spectator Expansion in multiple scattering theory.
Various other forms could also be found \cite{Sicil}.
Differences between one form or another reside primarily
in the treatment of  the many-body propagator $G_0(E)$. The Spectator
Expansion derives its name from the underlying idea that in lowest order
all target constituents but the initially struck one
(particle $i$) are `passive'.  In the next order all target
constituents but the $i$th and $j$th particle are
passive, and so on. In that sense, the Spectator Expansion resembles the
linked-cluster decomposition of nuclear structure \cite{Brandow}.

\subsection{The First Order Term}

\hspace*{10mm}
The first order term in the Spectator Expansion,
$\tau_i$ as given by Eq.~(\ref{eq:2.15}), is now examined.
Since for elastic scattering only $P\tau_i P$, or equivalently
$\langle\Phi_A| \tau_i | \Phi_A\rangle$ need be considered,
Eq.~(\ref{eq:2.15}) can be reexpressed with this in mind as
\begin{eqnarray}
\tau_i &=& v_{0i} + v_{0i}G_0(E) \tau_i - v_{0i} G_0(E) P \tau_i \nonumber \\
   &=& \hat{\tau_i} - \hat{\tau_i} G_0(E) P \tau_i, \label{eq:2.24}
\end{eqnarray}
or
\begin{equation}
\langle\Phi_A| \tau_i | \Phi_A\rangle = \langle\Phi_A| \hat{\tau_i}|
 \Phi_A\rangle - \langle\Phi_A| \hat{\tau_i}| \Phi_A\rangle \frac {1}
  {(E-E_A) - h_0 + i\varepsilon} \langle\Phi_A| \tau_i | \Phi_A\rangle,
 \label{eq:2.25}
\end{equation}
where $\hat{\tau_i}$ is defined as the solution of
\begin{equation}
\hat{\tau_i} = v_{0i} + v_{0i} G_0(E) \hat{\tau_i}. \label{eq:2.26}
\end{equation}
The combination of Eqs.~(\ref{eq:2.24}) and (\ref{eq:2.2}) corresponds
to the first order Watson scattering expansion \cite{Watson}. If the
projectile$\,-\,$target nucleon interaction is assumed to be the
same for all target nucleons and if isospin effects are neglected
then the KMT scattering integral equation \cite{KMT} can be derived from
the first order Watson scattering expansion.

\hspace*{10mm}
Since Eq.~(\ref{eq:2.25}) is a simple one-body integral equation,
the principal problem is to find a solution of Eq.~(\ref{eq:2.26}).
Of course, due to the many-body character
of $G_0(E)$, Eq.~(\ref{eq:2.26}) is a
many-body integral equation, and in fact no more easily solved
than the original equation Eq.~(\ref{eq:2.1}).
However, $G_0(E)$ may be written as
\begin{eqnarray}
G_0(E) &=& ( E - h_0 -H_A + i\varepsilon)^{-1}  \nonumber \\
    &=& ( E - h_0 -h_i -W_i -H^i + i\varepsilon)^{-1} \label{eq:2.27}
\end{eqnarray}
with
\begin{equation}
W_i = \sum_{j\neq i} V_{ij}  \label{eq:2.28}
\end{equation}
and
\begin{equation}
H^i = H_A -h_i -W_i.   \label{eq:2.29}
\end{equation}
Since $H^i$ has no explicit dependence on the $i$th particle, then
Eq.~(\ref{eq:2.26}) may be simplified by the replacement of $H^i$ by an
average energy $E^i$. This is not necessarily an approximation since
$G_0(E)$ might be regarded to be
\begin{equation}
G_0(E) = [(E-E^i) -h_0 -h_i -W_i - (H^i -E^i) + i\varepsilon]^{-1}
                    \label{eq:2.30}
\end{equation}
and $(H^i -E^i)$ could be set aside to be treated in the next order
of the expansion of the propagator $G_0(E)$. Thus, consider
now $G_0(E)$ to be $G_i(E)$, where
\begin{equation}
G_i(E)= [(E-E^i) -h_0 -h_i -W_i + i\varepsilon]^{-1}, \label{eq:2.31}
\end{equation}
so that $\hat{\tau_i} = \tilde{\tau_i} + ({\rm
higher~order~corrections})$, and Eq.~(\ref{eq:2.26}) reduces to
\begin{equation}
\tilde{\tau_i}=v_{0i} + v_{0i} G_i(E)\tilde{\tau_i}. \label{eq:2.32}
\end{equation}
Eq.~(\ref{eq:2.32}) can also be reexpressed as
\begin{equation}
\tilde{\tau_i} = t_{0i} + t_{0i} g_i W_i G_i(E) \tilde{\tau_i},
                         \label{eq:2.33}
\end{equation}
where the operators $t_{0i}$ and $g_i$ are defined to be
\begin{equation}
t_{0i} = v_{0i} + v_{0i} g_i t_{0i} \label{eq:2.34}
\end{equation}
and
\begin{equation}
g_i =[ (E-E^i) - h_0 -h_i + i\varepsilon]^{-1}. \label{eq:2.35}
\end{equation}

\hspace*{10mm}
The quantity $W_i$ represents the coupling of the struck target
nucleon to the residual nucleus. At this point, one could take the
attitude that a proper consideration of this quantity is not of first
order, and it should be put together with the next higher order in the
Spectator Expansion. In that case one would obtain the so-called
`$t^{free}$' or impulse approximation to the optical potential, which
can be viewed as $\hat{\tau_i} \approx \tilde{\tau_i} \approx t_{0i}$.
In the case of the impulse approximation, one never needs to solve
any integral equation for more than two particles. This has made the
impulse approximation very practical in
intermediate energy nuclear physics and has over many years led to
a large body of work being based upon this approximation \cite{LRay}.

\subsection{Coupling of the Struck Target Nucleon to the Nucleus}

\hspace*{10mm}
In the explicit treatment of the propagator $G_i(E)$ it is
necessary to consider
specific forms of the potential $W_i$, which represents the coupling of the
struck nucleon to the residual nucleus. In this paper $W_i$ is
treated as one-body operator, such as a shell-model or mean field
potential. The attitude is taken here that this potential is
already known and is extracted from single particle mean field
potentials as calculated in various studies of nuclear structure.
In this specific case, Eq.~(\ref{eq:2.33}) can be written as
\begin{equation}
\tilde{\tau_i} = t_{0i} + t_{0i} g_i {\cal T}_i g_i \tilde{\tau_i},
                       \label{eq:2.36}
\end{equation}
with ${\cal T}_i$ being given as the solution of a Lippmann-Schwinger
type equation with the potential $W_i$ as the driving term
\begin{equation}
{\cal T}_i = W_i + W_i g_i {\cal T}_i.  \label{eq:2.37}
\end{equation}

This is the approach taken in the calculations presented in this
paper as well as in earlier work \cite{med1,med2,med3}. While
Eqs.~(\ref{eq:2.36}, \ref{eq:2.37}) are completely equivalent to
Eq.~(\ref{eq:2.33}), a justification for the substitution
of a Hartree-Fock or any other single particle mean field potential
taken from a nuclear structure calculation is not strictly within
the theoretical prerequisites of the Spectator Expansion, which
demands that all of the two-body interactions be consistently
represented by $v_{0i}$.  Standard mean field or shell model
calculations use an effective NN interaction for the reason that
present microscopic nuclear structure calculations are unable
simultaneously to use realistic free NN potentials and predict
the experimental results.  Hence it is not physically unreasonable
to substitute a mean field potential for $W_i$, but
this choice is {\it defacto} outside the strict demands of the Spectator
Expansion.

Using the expression given in Eq.~(\ref{eq:2.28}), for $W_i$,
Eq.~(\ref{eq:2.33}) can be reformulated as
\begin{eqnarray}
\tilde{\tau_i} &=& t_{0i} + t_{0i} g_i \sum_{j\neq i} v_{ij}
  \left[ \frac {1}{g_i^{-1} - \sum_{k\neq i} v_{ik}+i\varepsilon }
   \right]  \tilde{\tau_i} \nonumber \\
 & = & t_{0i} + t_{0i} g_i \sum_{j\neq i} v_{ij}
  \left[ \frac {1}{g_i^{-1} - v_{ij} +i\varepsilon}  \right.
                                 \nonumber \\
 & ~ & + \left.  \sum_{j\neq i} v_{ij}\frac {1}
{g_i^{-1} - v_{ij} +i\varepsilon} \sum_{k\neq i,j} v_{ik} \frac {1}
  {g_i^{-1} - \sum_{l\neq i} v_{il} +i\varepsilon} \right] \tilde{\tau_i}
                                 \nonumber \\
 &=& \bar{\tau_i} + \bar{\tau_i} g_i \sum_{j\neq i}
v_{ij} \frac {1}{g_i^{-1} - v_{ij} +i\varepsilon}
\sum_{k\neq i,j} v_{ik} \frac {1}{g_i^{-1} - \sum_{l\neq i} v_{il}
+i\varepsilon} \tilde{\tau_i}, \label{eq:2.38}
\end{eqnarray}
where
\begin{equation}
\bar{\tau_i}  =  t_{0i} + t_{0i} g_i \sum_{j\neq i} t_{ij} g_i
\bar{\tau_i}, \label{eq:2.39}
\end{equation}
and
\begin{equation}
 t_{ij}  =  v_{ij} + v_{ij} g_i t_{ij} .
\end{equation}

\hspace*{10mm}
Since the last term in Eq.~(\ref{eq:2.38}) always involves at least
three different target particles ($i,j,k$), this term is of higher
order and is safely neglected at present.
Thus the operator $\tilde{\tau_i}$ can be written as
\begin{eqnarray}
\tilde{\tau_i} &=& \bar{\tau_i} + \cdots   \nonumber \\
   &=& t_{0i} + \sum_{j\neq i} \eta_{ij} + \cdots \label{eq:2.40}
\end{eqnarray}
where
\begin{eqnarray}
 \sum_{j\neq i} \eta_{ij} & \equiv & \bar{\tau_i} - t_{0i} \nonumber \\
\eta_{ij} & = & t_{0i} g_i t_{ij} g_i t_{0i} + t_{0i} g_i t_{ij} g_i
\eta_{ij},  \label{eq:2.41}
\end{eqnarray}
where $\eta_{ij}$ neglects the terms involving three target nucleons
that arise from Eq.~(\ref{eq:2.38}).
This treatment of the interaction of the struck target nucleon with
the residual nucleus, though more complicated, is completely
consistent with the spirit of the Spectator Expansion.

\hspace*{10mm}
The term $\sum_{j\neq i} \eta_{ij}$ involves two active target particles
and thus represents a second order Spectator Expansion correction to the
first order term considered in this paper. In fact, it could equally well
be considered together with the second order term $\tau_{ij}$ shown in
Eq.~(\ref{eq:2.17}). We have found it expedient, however, to define the
Spectator Expansion as given in Eq.~(\ref{eq:2.17}).  Since this
expansion is performed in terms of quantities which in themselves
contain many-body propagators, each of the ingredients,
$\tilde{\tau}_i$, $\tilde{\tau}_{ij}$, etc. may themselves be
expanded in a spectator expansion.  This amounts to expanding the
many-body propagator also according to the number of active participants.
Another reason to distinguish the corrections to the
propagator and the explicit second order term is that the second order terms in
Eq.~(\ref{eq:2.17}), correspond to contributions which arise from the
$Q$ space, whereas the second term in Eq.~(\ref{eq:2.40}) remains in the
$P$ elastic space at the first order level.

\subsection{The Second Order Term}
Since second order corrections in the propagator should at least in
principle be considered simultaneously with the second order
corrections in the multiple scattering expansion, the second term
in Eq.~(\ref{eq:2.17}) is examined in detail.   This term may be
written as
\begin{equation}
\sum_{i,j\neq i} \tau_{ij} = \sum_{i>j}(\tilde{U}_{ij} - \tau_i -
      \tau_j).   \label{eq:2.42}
\end{equation}
The subtraction of the two-body contribution in
Eq.~(\ref{eq:2.42}) plays an important role in that any double
counting of the two-body term is removed.  This also enables us to
see explicitly the three-body nature of the second order term.
We start from Eq.(\ref{eq:2.23}), which can be expressed as
\begin{eqnarray}
\tilde{U}_{ij}&=&(v_{0i}+v_{0j})+(v_{0i}+v_{0j}) G_0(E) \tilde{U}_{ij}
      - (v_{0i}+v_{0j})G_0(E) P \tilde{U}_{ij} \nonumber \\
 &=& \hat{t}_{0ij} - \hat{t}_{0ij} G_0(E) P \tilde{U}_{ij}, \label{eq:2.43}
\end{eqnarray}
where $\hat{t}_{0ij}$ is defined as
\begin{equation}
\hat{t}_{0ij} = (v_{0i}+v_{0j})+(v_{0i}+v_{0j}) G_0(E) \hat{t}_{0ij}.
                           \label{eq:2.44}
\end{equation}
Thus $\langle\Phi_A|\tilde{U}_{ij}|\Phi_A\rangle$ is obtained from
$\langle\Phi_A| \hat{t}_{0ij}|\Phi_A\rangle$ through the solution of a
straightforward one-body integral equation in a way similar to the
manner $\langle\Phi_A| \tau_i |\Phi_A\rangle$ is obtained from
$\langle\Phi_A|\hat{\tau_i}|\Phi_A\rangle$.  In this case
Eq.~(\ref{eq:2.44}) is a full three-body equation.

\hspace*{10mm}
Once again notice that the propagator $G_0(E)$ in Eq.~(\ref{eq:2.44}) is a
complicated many-body operator. Consistent with the spirit of the
Spectator Expansion the propagator can be written as
\begin{equation}
G_0(E) = \left[ (E-h_0 -H^{ij} + i\varepsilon) -h_i -h_j - v_{ij} -
\sum_{k\neq i,j} (v_{ik}+v_{jk})\right] ^{-1} \label{eq:2.45}
\end{equation}
and a three particle Greens function is defined to be
\begin{equation}
G_{ij}(E)= \left[ (E- E^{ij} -h_0 + i\varepsilon) -h_i -h_j - v_{ij}
  \right] ^{-1}.
                     \label{eq:2.46}
\end{equation}
Using the same procedure used in the first order term,
$G_{ij}(E)$ is substituted for $G_0(E)$ in Eq.~(\ref{eq:2.44}) to obtain
\begin{equation}
\tilde{t}_{0ij} = (v_{0i}+v_{0j})+(v_{0i}+v_{0j}) G_{ij} \tilde{t}_{0ij},
                 \label{eq:2.47}
\end{equation}
where the effective 3-body t-matrix becomes
\begin{equation}
\hat{t}_{0ij} = \tilde{t}_{0ij} + \cdots .    \label{eq:2.48}
\end{equation}
The truncation of the propagator $G_0(E)$ from Eq.~(\ref{eq:2.45}) to the
form given in Eq.~(\ref{eq:2.46}) is once again tantamount to relegating
the coupling of the active target nucleons to the next higher order
term in the expansion of the propagator.

\hspace*{10mm}
Actual calculations of the three-body corrections to the first order
optical potential as given in Eq.~(\ref{eq:2.41}) and Eq.~(\ref{eq:2.47})
are extraordinarily difficult without further approximations.
In this paper we do not attempt to calculate these higher order
contributions to the Spectator Expansion. But for the sake of
conceptual clarity the propagator corrections in first order, as
presented in this work, should be seen in the context of the next higher
order of the Spectator Expansion, since both are given through
three-body type equations.

\section{Results and Discussion}

\subsection{Details of the Calculation}

\hspace*{10mm}
In this paper the study of the elastic scattering of neutrons and
protons from spin zero target nuclei at energies that range from
65 and 400~MeV (incident projectile energy) is strictly first order in the
Spectator Expansion. Here the correction to the propagator
$G_{0}(E)$ due to the coupling of the initially struck target nucleon
to the residual target is considered to be first order. As
outlined in subsection C of the previous section this calculation
includes the modification of the free propagator due to
the `nuclear medium'.  The operator
${\cal T}_i$, representing the scattering of the struck target particle
$i$ from the residual nucleus, is calculated through the use of a
one-body potential $W_{i}$.  Nonlocal, spin-dependent potentials
derived from realistic nuclear mean field models are used
to represent the potential $W_{i}$ given in
Eq.~(\ref{eq:2.37}). Two different mean field potentials are used in
these calculations in order to isolate any model dependence which may
exist. One is the nonrelativistic, non-local mean field
potential taken from a
Hartree-Fock-Bogolyubov microscopic nuclear structure calculation,
which utilizes the density-dependent finite-ranged {\it Gogny D1S}
nucleon-nucleon interaction {\cite {HFB,Gogny}}. This model has been shown
to provide accurate descriptions of a variety of nuclear structure
effects. Calculations using this potential as $W_{i}$ will be referred
to as HFB. The second choice involves a nonrelativistic, local
reduction of the
mean field potential resulting from a Dirac-Hartree calculation
based upon the $\sigma-\omega$ model \cite{DH}. The calculations
with this potential will be referred to as DH. Comparisons of
calculations with these two models may serve to indicate the
sensitivity of the elastic scattering predictions to the model of
the nuclear mean field potential. The results suggest that there
is a slight sensitivity to the choice of the mean field potential,
however this `uncertainty' is smaller than the overall size of the
medium correction. One might therefore expect that any
reasonable model of this kind, which describes nuclear structure
could give qualitatively similar results. A step by step description
of the implementation of the nuclear mean field potential,
consistent with
the framework of the Spectator Expansion is given in Ref.~\cite {med1}.

\hspace*{10mm}
The treatment of the propagator modification through a nuclear
mean field potential taken from structure calculations, although
a valid approach, may not be completely satisfactory.
In keeping full consistency with the theory of multiple scattering,
it may be better to treat the operator ${\cal T}_{i}$ as
\begin{equation}
{\cal T}_{i} = \sum_{j\neq i}t_{ij} + \cdots \label{eq:3.1}
\end{equation}
as outlined in subsection C of the previous section
of Eqs.~(\ref{eq:2.40}) and (\ref{eq:2.41}), where $t_{ij}$ is defined
in Eq.~(\ref{eq:2.39})].
Calculations based on this approach are much more difficult than any
performed so far, and though not intrinsically intractable, have
been postponed. The structure of Eqs.~(\ref{eq:2.40}) and (\ref{eq:2.41})
is very
similar to the calculation of the second order in the Spectator Expansion as
outlined in subsection D of the previous section, and both should
be treated in the same order and in a similar manner. Further,
note that an approximate treatment of the three-body kinematics
involving the scattering of the struck target nucleon from the
residual nucleus is used and is discussed in length in Ref.~\cite{med1}.

\hspace*{10mm}
The nucleon-nucleon (NN) t-matrix is another crucial ingredient
in these calculations. The quality and extensiveness of the
nucleon-nucleus observables we attempt to predict require
trustworthy representations of the NN interaction. For
convenience the calculations presented here use the free NN
interaction based upon the full Bonn potential\cite{Bonn}, giving
$t_{oi}=t^{free}_{oi}$. This interaction includes the effects
of relativistic kinematics, retarded meson propagators as given
by time-ordered  perturbation theory, and crossed and iterative meson
exchanges with NN, N$\Delta$, and $\Delta\Delta$ intermediate states.
A comparison of nucleon-nucleus observables based on different models
for the  NN interaction is deferred to a later time. It should be clearly
stated that even if the underlying models for the NN interaction
accurately describe the `on-shell' NN data, there may still exist
`off-shell' differences between the various models, which could
affect the predictions of the elastic nucleon-nucleus observables.

\hspace*{10mm}
The first order folded effective NN t-matrix is then constructed
with the operator, $\tilde{\tau}_i$, from Eq.~(\ref{eq:2.36}):
\begin{equation}
\langle \tilde{\tau}_{eff}\rangle = \langle{\vec k'}_0\Psi_A|
   \sum_i \tilde{\tau}_i
  |{\vec k}_0\Psi_A\rangle ~. \label{eq:3.2}
\end{equation}
These calculations are performed in momentum space and include
spin degrees of freedom.  The first order optical potential is
then evaluated by solving Eq.~(\ref{eq:2.24}) in the folded form.
In the present calculations, which are performed in momentum space,
$\langle \tilde{\tau}_{eff}\rangle$ enters in the `optimum
factorized' or `off-shell $\tau\rho$' form \cite{ernst,pttw} as
\begin{equation}
\langle \tilde{\tau}_{eff}\rangle \approx \tilde{\tau}(q,{\cal K};E)
\rho(q) ~, \label{eq:3.3}
\end{equation}
where ${\vec q}={\vec k}_0'-{\vec k}_0$ and
${\vec {\cal K}}=\frac{1}{2}\left({\vec k}'_0+{\vec k}_0\right)$;
${\vec k}_0'$ and ${\vec k}_0$ are the final and initial momenta
of the projectile.
This corresponds to a  steepest descent evaluation of the
`full-folding' integral, in which the non-local operator
$\tilde{\tau}$ is convoluted with the
density $\rho(q)$ as indicated schematically in
Eq.~(\ref{eq:3.2}). For harmonic oscillator model densities it has
been shown that the optimum factorized form represents the nonlocal
character of $U_{opt}$ qualitatively in the intermediate energy
regime \cite{FF,FFC}. Complete `full-folding' calculations with
more realistic nuclear densities are in progress.  It is to be
understood that we perform all spin summations in obtaining
$U_{opt}$.  This reduces the required NN t-matrix elements to
a spin-independent component (corresponding to the Wolfenstein
amplitude $A$) and a spin-orbit component (corresponding to
the Wolfenstein amplitude $C$).
All scattering calculations presented here contain an additional factor
in the optical potential to account for the transformation of the NN
t-matrix from the two-nucleon c.m. frame to the nucleon-nucleus c.m.
frame \cite{pttw}.

\hspace*{10mm}
Another uncertainty in the present calculations lies in the lack
of completely reliable target wave functions of the accuracy
required. The best guide for the distribution of matter in
nuclei is the information extracted from electron scattering
\cite{electron}.  This information gives a reasonably good
picture of the average single particle proton density
especially about the surface, and it is used in the calculations
presented in this paper to represent the proton densities. The neutron
densities used are those taken from the Hartree-Fock-Bogolyubov
calculation described above \cite{HFB}. In principle, it
would appear more consistent to employ the proton densities
obtained from the same calculation. However, there are small
differences between the calculated proton densities and the
measured ones. These small differences are, however, large enough
to influence slightly the predictions of the nucleon-nucleus
observables.  Therefore, the measured proton distributions
were thought to be more reliable and are used throughout the
present calculations. This does leave
a question about the reliability of the calculated
neutron distributions which were used in the present calculations.
A study on the sensitivity of the proton-nucleus observables
to slight variations in the neutron distributions was presented
in Ref.~\cite {ISO}.

\hspace*{10mm}
For the proton-nucleus scattering calculations the Coulomb
interaction between the projectile and the target is included
using the exact formalism described in Ref.~\cite{coul}.
Although the multiple scattering calculations are performed fully
in momentum space, so as to be able to include easily nonlocal and
off-shell effects, the point Coulomb contributions are described
by using Coulomb scattering wave functions in coordinate space.
There are no cut-off parameters necessary in this technique.

\subsection{Total Cross Section for Neutron Scattering}

\hspace*{10mm}
In nuclear structure calculations the binding energy of the system in
its ground state together with energies of certain low lying excited
states are the experimental information which must be closely reproduced
to establish the reliability of the model wave functions
and the various physical matrix elements implied thereby.
In the present calculations of elastic nucleon-nucleus scattering
the neutron total cross section, as a function of
scattering energy, should serve as a similar figure of merit.
In the early 1990's extensive high precision measurements of neutron
total cross sections became available for a variety of target nuclei
\cite{ndata,Roger} and can now be used to discriminate between scattering
calculations in the above indicated fashion.

\hspace*{10mm}
In Fig. 1 total neutron cross section data for $^{12}$C, $^{16}$O,
$^{28}$Si, $^{40}$Ca, $^{90}$Zr and $^{208}$Pb are shown along
with various calculations of $\sigma _{tot}(E)$ at a number of
energies. Because the data are so extensive, the `usual' procedure
has been reversed and the data are represented by dotted
curves. The `jitter' in these curves may be taken as indicative of  the
experimental uncertainty. The statistical error bars themselves are of
the order of 1\% and could not be distinguished in the
figures. The discrete
points correspond to the calculated results. The solid diamonds represent
the calculations as described in Section~A and include the modification of
the free propagator through the Hartree-Fock-Bogolyubov mean
field \cite{HFB}, except for $^{208}$Pb where $W_i$ is taken from
the DH case. In each case the predictions are in accord
with the data from $\rsim 65$~MeV for the light nuclei and $\rsim 100$~MeV
for the heavier nuclei. That is, the theoretical predictions do
extremely well in predicting the
energy dependence of the total neutron cross section beyond the point
where the data exhibit a pronounced structure.

\hspace*{10mm}
Since a detailed discussion on the description of neutron total cross
sections for $^{16}$O and $^{40}$Ca was recently published \cite{med2}
comments on this subject will be somewhat restricted. As an
indication as to how much has been gained by eliminating necessary earlier
approximations to the full first order theory, points, represented
as crosses, are shown at 100 and 200 MeV for  $^{16}$O, $^{40}$Ca,
$^{90}$Zr and $^{208}$Pb which were calculated using the so-called `local
free $\tau \rho$' approximation. This consists of multiplying the
on-shell NN scattering amplitude $t(q)$ with the one nucleon density
$\rho(q)$ for the target nucleus. For many years this simple approximation
was taken to represent the first order theory (of course, it was not
possible to perform more difficult calculations at that time).
The total cross sections calculated this way are invariably above
the full calculation represented by the solid diamonds. These
`local free $\tau\rho$' results are also significantly larger then
the experimental values, in some cases, especially for the heavier
elements, this discrepancy can be as large as $25-30\%$. This gross
failure of the local approximation casts serious doubt upon some of
its early successes, and certainly creates serious reservations
about the many attempts to account for the failures of the local
approximation by the introduction of new effects, which are not
cleanly consistent with a many-body scattering theory.

\hspace*{10mm}
Of greater current interest is the
difference between the points represented as stars and the solid
diamonds. The stars are calculations performed with the free
propagator $G_{0}(E)$ of Eq.~(\ref{eq:2.27}), where the target
Hamiltonian $H_A$ is approximated by a c-number.
Therefore  these points contain the complete off-shell structure
of the NN t-matrix, but neglect the coupling of the struck target
nucleon to the residual nucleus. The difference between the stars
and the solid diamonds represent the size of this effect, i.e.
the coupling of the struck target nucleon to the residual nucleus.
As Fig.~1 shows, and as expected, the absolute size of the effect
grows as the nuclei become heavier. In addition it is most prevalent
in the regime between 100 and 200~MeV projectile energy and becomes
almost negligible at higher energies.  For all nuclei under
consideration at
$300$ and $400$~MeV the propagator modification has no discernible
effect.  It is most satisfying to observe that whenever this
correction is significant, it moves the calculated results closer
to the measurements.

\subsection{Proton Elastic Scattering Observables}

\hspace*{10mm}
Obviously there are no comparable total cross section data for proton
scattering. On the other hand, there is relatively little
experimental information on elastic angular distributions and no
spin-observables for neutron scattering from nuclei.  Thus for a
more detailed look at nucleon-nucleus elastic scattering, the
scattering of protons from nuclei is examined.   For the proton
case the Coulomb interaction between the projectile and the target
is included using the exact method developed in Refs.~\cite{coul}.
In a recent paper \cite{med3} elastic proton  and neutron scattering
observables at 65~MeV projectile energy for $^{12}$C, $^{16}$O,
$^{28}$Si, $^{40}$Ca, $^{90}$Zr and $^{208}$Pb were shown. These
calculations were performed at an energy which was considered by
many to be below the regime of applicability of the first order
Spectator Expansion. However there is a wealth of experimental data
at this energy for the above mentioned nuclei.  This data includes
the differential cross section $\frac{d\sigma}{d\Omega}$ , the
analyzing power $A_y$ and the spin rotation function $Q$ (with the
exception that for $^{28}$Si and $^{56}$Fe there are no measurements
of $Q$). At this low energy fairly good agreement between the
predictions and the measurements was observed. It was very clear
that the inclusion of the coupling of the struck target nucleon to
the residual nucleus considerably improved the description of the
data. This was especially true for the description of the spin
rotation parameter $Q$, in which case the improvement was dramatic.
In this paper further calculations at other energies in the regime
between 80 and 300~MeV are presented.  Elastic proton scattering
observables are calculated for a variety of light as well as heavy
spin-zero targets at a variety of energies.  Here predictions are
presented only for those targets and energies where spin observable
data exist.  Unfortunately in the regime below 200~MeV,
there is no other energy like 65~MeV
where the proton spin observables have been measured for many nuclei
in a similarly systematic manner.

\hspace*{10mm}
The calculations chosen here are limited to energies
below pion-production threshold.
The reason is twofold. First, the NN t-matrix, which represents one
of the critical inputs to the calculations is not as well established
at higher energies. Second, the correction to the free propagator
due to the presence of the nuclear medium, which is the new ingredient
in these calculations decreases in importance as the energy increases.
This is shown for the total cross section in Fig.~1. In Figs. 2-14
the angular distribution, the analyzing power and the spin rotation
function for elastic proton scattering are shown for $^{12}$C at 200 MeV;
$^{16}$O at 100, 200, and 318~MeV, $^{28}$Si at 80, 135 and 200~MeV,
$^{40}$Ca at 80~MeV, $^{90}$Zr at 65, 80 and 160~MeV, and $^{208}$Pb at
80 and 200~MeV, respectively.

\hspace*{10mm}
To show the dependence on the choice of the mean field potential
$W_{i}$ (Eqs.~(\ref{eq:2.31}) and (\ref{eq:2.33})) calculations
with two different mean field potentials representing the
operator $W_i$ are displayed. The solid line corresponds to the
results based upon the HFB potential \cite{HFB}, whereas the
dashed curve is based upon the DH potential \cite{DH}. For the
calculations of proton scattering from $^{208}$Pb the DH mean
field potential only is used (solid line), since the choice and
size of the basis function representing $W_{i}$ was not adequate
for the HFB mean field potential in this case.  Calculations,
where no medium contributions are included, correspond to the
free `off-shell $t\rho$' approximation and are given by the
dash-dotted line.

\hspace*{10mm}
The figures show that both calculations which incorporate the
coupling  of the struck nucleon to the residual nucleus provide a good
representation of the data except at very large scattering angles.
That this description, while very good, is not perfect is easily
understandable since various corrections to the many body theory
still remain unexplored. The calculations are limited to lowest
order in the spectator expansion, and are carried out in the
optimum factorized form, which takes only the non-local structure
of the NN t-matrix into account.  Finally, the full three-body
structure involved in the coupling of the struck nucleon to the
residual nucleus has not yet been considered.  In addition, for large
angles the Pauli exchange of the Coulomb term could play
a role. Those effects are not included in the present calculations.

\hspace*{10mm}
For the lighter target nuclei in Figs.~2-8 the correction to the
propagator causes the diffraction minima in the predictions to
move slightly to higher angles and be closer to the data.  The
improvement of the theoretical predictions is more apparent in the
spin observables, especially in Fig.~6.
At lower energies (Figs. 3, 6, and 9) the propagator modifications
causes a characteristic shift of the spin rotation function $Q$.
Unfortunately there are no measurements of $Q$ at the energies
presented here. In an earlier work \cite{med3},
presenting only calculations at
65~MeV projectile energy, the propagator modification brought the
the calculations in excellent agreement with the measured values for
$Q$. So we expect that measurements of $Q$ around 100~MeV would also be
close to our calculations.
At projectile kinetic energies of 200~MeV or above the
full calculations provide very good results with respect to the
data. The modification of the propagator effect only
$A_y$ in a very moderate fashion.

\hspace*{10mm}
For the heavier nuclei in Figs.~10-14 one would expect the effects
of the medium modifying the propagator to be more pronounced, which
is indeed the case. The shift of the diffraction minima to larger angles
is clearly visible for the heavy nuclei $^{90}$Zr and $^{208}$Pb,
especially at lower energies (Fig.~10, 11, and 12).
For $^{90}$Zr at 65~MeV (Fig.~10) the propagator modification
has a significant effect on the observables. The minima in the
diffraction pattern of the differential cross section coincide with the
measured ones, indicating a correct prediction of the size of
the nucleus. Furthermore, the overall size of
$d\sigma/d\Omega$ is predicted correctly over about five orders
of magnitude. The effect on the spin observables is equally dramatic and
causes the predictions of  $A_y$ and $Q$  to agree remarkably well with
the data. We included Fig.~10, though already being published in
Ref.~\cite{med3}, since at lower energies the only measurements of
$Q$ exist at 65~MeV.
Again, for $^{90}$Zr at 80~MeV (Fig.~11) the
propagator modification has a significant effect on the differential
cross section and the spin observables
and causes the predictions of $A_y$ to agree very well with the
data. A similar tendency can be seen for
$^{208}$Pb at the same energy (Fig.~13).
At 160~MeV the medium effects are not as pronounced  for $^{90}$Zr
except at larger angles where the full calculation provides
better agreement with the data (Fig.~12).
For $^{208}$Pb at 200~MeV in Fig.~14 in the spin
rotation there are differences caused by the modification of the
propagator, but the effects on the predictions of the data are
unclear.

\hspace*{10mm}
In order to indicate the progress which has been made in the
calculation of proton-nucleus elastic scattering, results for
selected cases obtained with the local, `on-shell $t\rho$'
approximation are shown, where the off-shell contributions of
both, the NN t-matrix as well as the density matrix are neglected.
Those `local' calculations are represented by the dotted lines in
Figs.~3, 4, and 14 for $^{16}$O at 100 and 200~MeV and for $^{208}$Pb
at 200~MeV, respectively. As was already the case for the description of
neutron total cross sections, the local calculations show moderate
deficiencies for a light nucleus like $^{16}$O, manifested in an
overprediction of the differential cross section, a lack of
structure in the spin observable $A_y$, and differences in the spin
rotation function $Q$. For a heavy nucleus like $^{208}$Pb the local
calculation looks disastrous, leading already to a severe
overprediction of the size  of the nucleus as seen in the diffraction
pattern of the differential cross section.  This lack of agreement
led in the past to the study of corrections to the local
approximation specifically in heavy nuclei \cite{LRay}.
It is a very satisfactory result for us to find that the first order
calculation in the Spectator Expansion together with a consistent
treatment of the propagator modification leads to a good description of
the elastic proton-nucleus observables for light as well as heavy nuclei
in the energy regime especially between $\sim$65 and 400 MeV projectile
energy.

\section{Conclusion}

\hspace*{10mm}
The Spectator Expansion of Multiple Scattering Theory is described in
detail.  The optical potential is expanded into a series
predicated upon the idea that the dominant effect is the two-body
interaction between the projectile and one of the nucleons in the
target.  The number of target nucleons interacting directly with
the projectile determines the ordering of the scattering series.
Complexities due to the free many-body
propagator for the projectile$\,-\,$target system also play a
significant role and are treated within a consistent theoretical
framework within the Spectator Expansion.  The first order theory and
the treatment of the many-body propagator due to effects from the
residual nucleus are presented, along with a formal description of
the second-order contribution.

\hspace*{10mm}
Predictions from rigorous calculations of elastic nucleon-nucleus
scattering at projectile kinetic energies in the range $\sim$65 to 400~MeV
provide excellent agreement with the experimental data.  In this
case the basic inputs to the calculation are the free fully
off-shell NN interactions and
realistic nuclear densities.  Modifications to the propagator were
calculated using static potentials taken from microscopic mean field
structure calculations.  It is found that as the
calculations include more complex degrees of freedom within a
well-defined theoretical framework, the predictions invariably
provide an improved description of the data.
The first order Spectator Expansion provides an excellent
{\it {a priori}} description of the extensive data for nucleon-nucleus
scattering data from $\sim$65 to 400~MeV for modest momentum transfers.
These results are in fact good enough to encourage speculation that
further work may soon yield new information about neutron distributions and
nuclear correlations in nuclei.

\vspace{5mm}

\vfill
\acknowledgments
The computational support of the the Ohio Supercomputer Center
under Grants No.~PHS206 and PDS150 is gratefully acknowledged.
This work was performed in part under the auspices of the U.~S.
Department of Energy under contracts No. DE-FG02-93ER40756 with
Ohio University, DE-AC05-84OR21400 with Martin Marietta Energy
Systems, Inc., and DE-FG05-87ER40376 with Vanderbilt University.
This research has also been supported in part by the U.S.
Department of Energy, Office of Scientific Computing under the
High Performance Computing and Communications Program (HPCC) as
a Grand Challenge project titled `the Quantum Structure of Matter'.

%----------------------------------------------------

\pagebreak

%%%%%%%%%%%%%%%%%%%%%%%%%%%%%%%%%%%%%%%%%%%%%%%%%%%%%%

\pagebreak
%%%%%%%%%%%%%%%%%%%%%%%%%%%%%%%%%%%%%%%%%%%%%%%%%%%%%%
\noindent
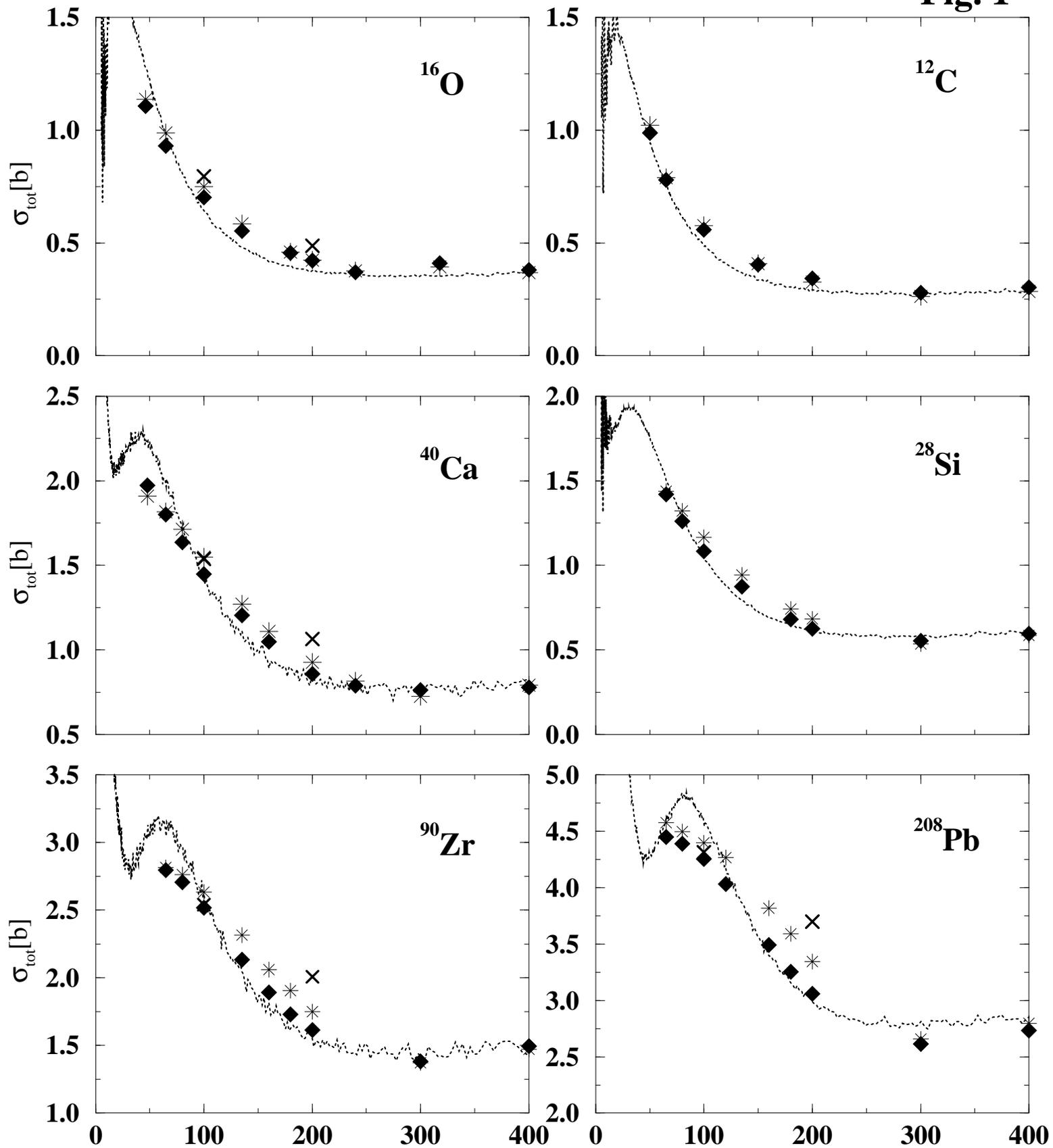
\begin{figure}
\caption{The neutron-nucleus  total cross-sections for
         scattering from $^{12}$C, $^{16}$O, $^{28}$Si, $^{40}$Ca,
         $^{90}$Zr, and $^{208}$Pb are shown as
         a function of the incident neutron kinetic energy.
         The dotted line represents the data taken from
         Ref.~\protect\cite{ndata,Roger}. The solid diamonds correspond
         to the calculations including the propagator modification due
         to the HFB mean field \protect\cite{HFB} (in case of
         $^{208}$Pb of the DH mean field \protect\cite{DH}).  The star
         symbols indicate the `free' calculations using the full Bonn
         free NN t-matrix \protect\cite{Bonn} only. The cross symbols
         represent a local `on-shell $t\rho$' calculation, which uses
         only  the on-shell values of the same t-matrix. \label{fig1}}
\end{figure}

\noindent
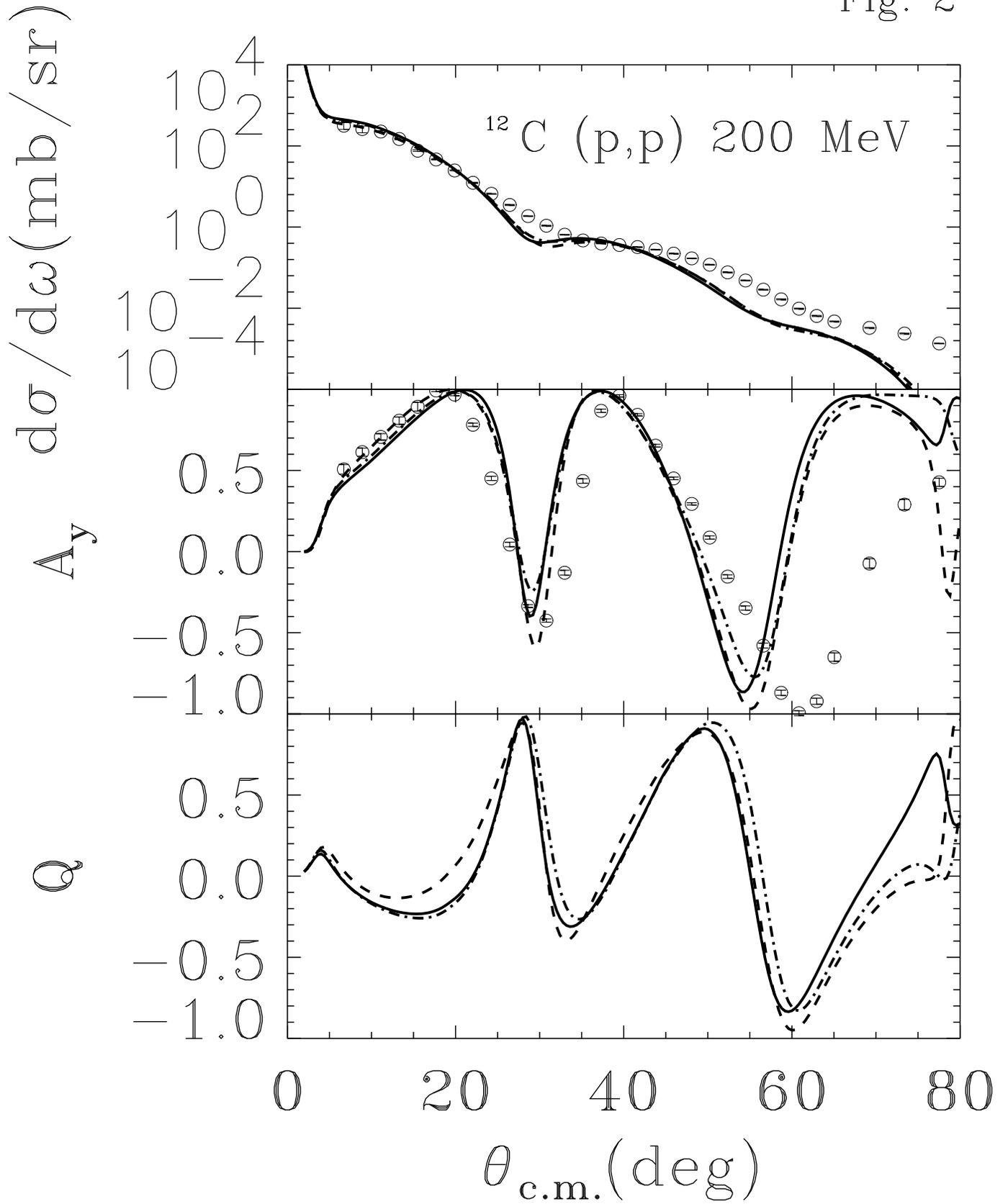
\begin{figure}
\caption{ The angular distribution of the differential cross-section
         ($\frac{d\sigma}{d\Omega}$), analyzing power ($A_y$) and
         spin rotation function ($Q$) are shown for elastic proton
         scattering from $^{12}$C at 200 MeV laboratory energy.  All
         calculations are performed with a first-order optical potential
         obtained from the full Bonn interaction \protect\cite{Bonn}
         in the optimum factorized form.  The solid curve includes
         the modification of the propagator due to the HFB mean
         field \protect\cite{HFB}, the dashed curve the one due to the
         DH mean field \protect\cite{DH}.  The free impulse approximation
         is given by the dash-dotted curve.  The data are taken from
         Ref.~\protect\cite{C200}.\label{fig2} }
\end{figure}

\noindent
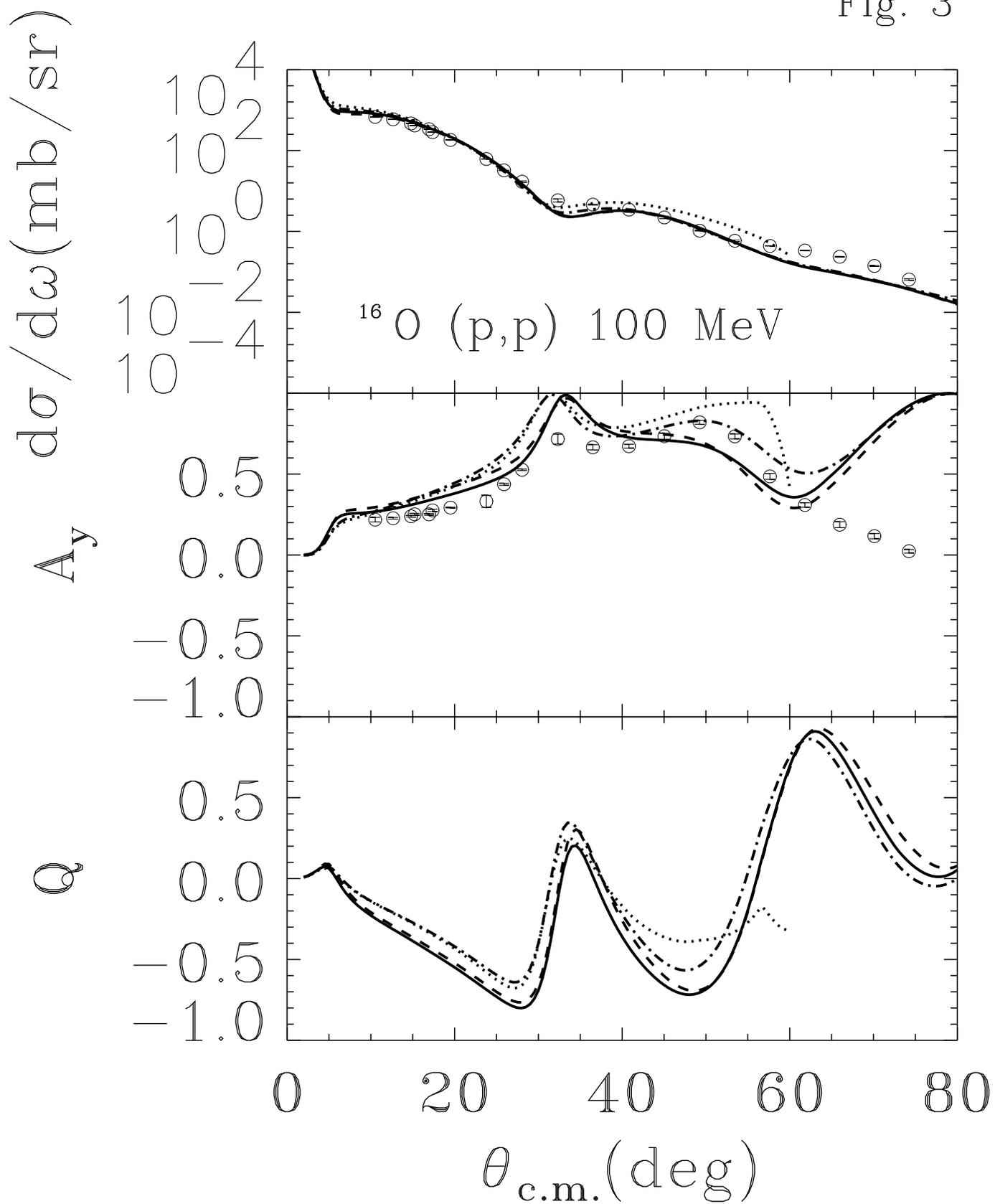
\begin{figure}
\caption{Same as Fig.~2,  except for $^{16}$O at 100 MeV proton
         kinetic energy. The dotted line represents a local
         `on-shell $t\rho$' calculation which uses only the
         on-shell values of the same t-matrix. The data are from
         Ref.~\protect\cite{O100}. \label{fig3}}
\end{figure}

\noindent
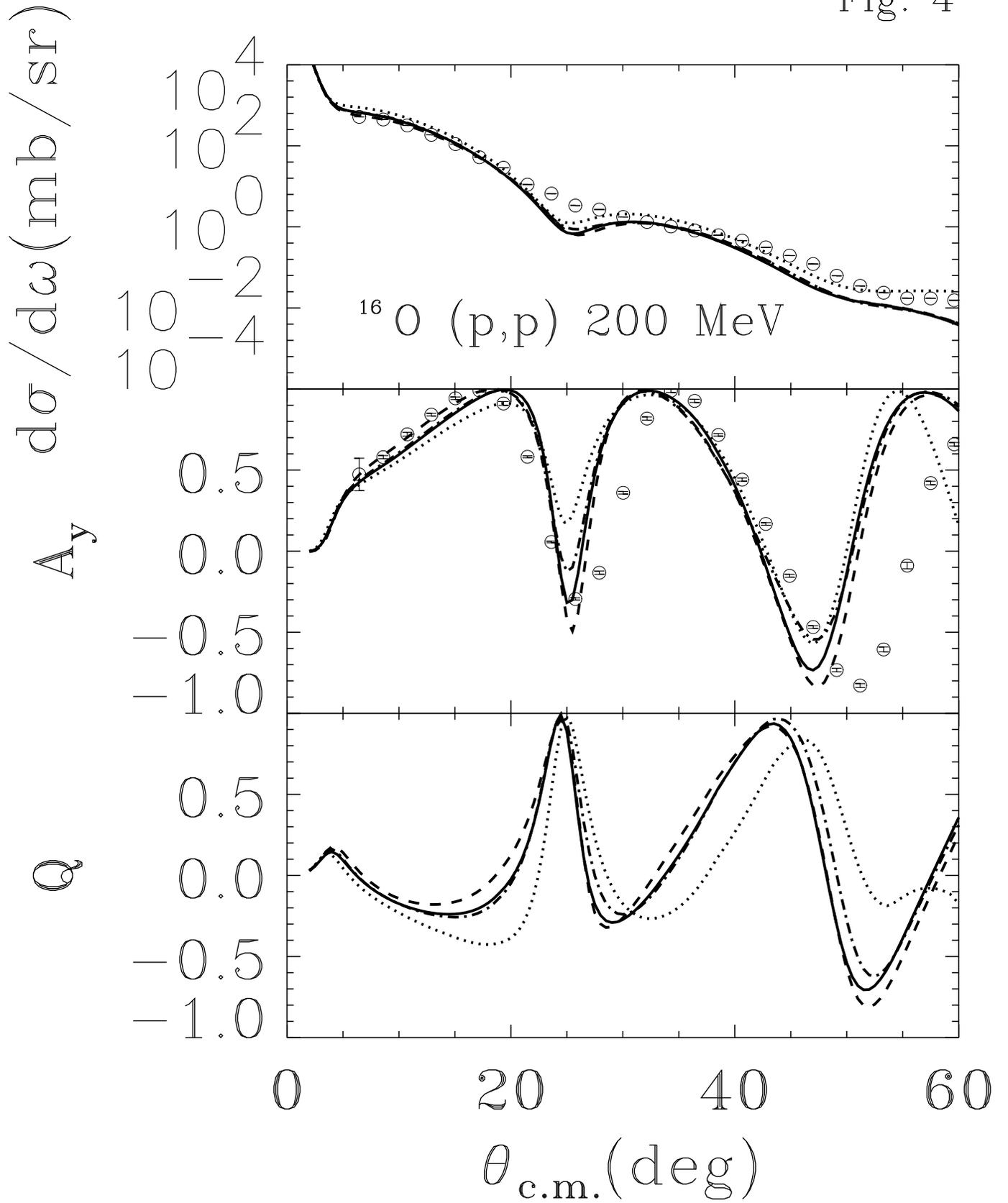
\begin{figure}
\caption{  Same as Fig.~3,  except the projectile kinetic energy is
200~MeV.  The dotted line represents a local
`on-shell $t\rho$' calculation which uses only the
 on-shell values of the same t-matrix.
The data are from Ref. \protect\cite{O200}.\label{fig4} }
\end{figure}

\noindent
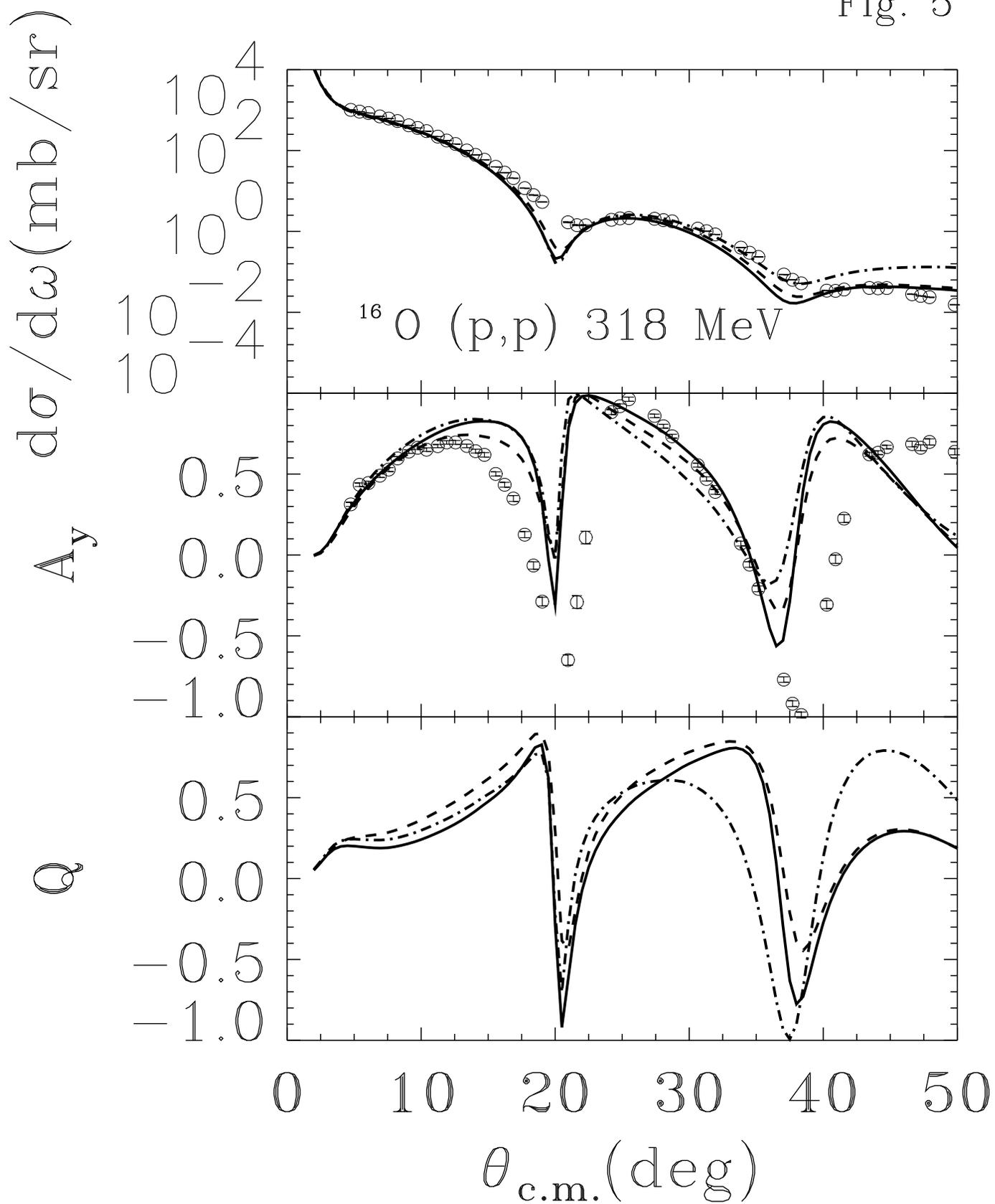
\begin{figure}
\caption{  Same as Fig.~3,  except the projectile kinetic energy is
318~MeV, and the data are from Ref. \protect\cite{O318}. \label{fig5} }
\end{figure}

\noindent
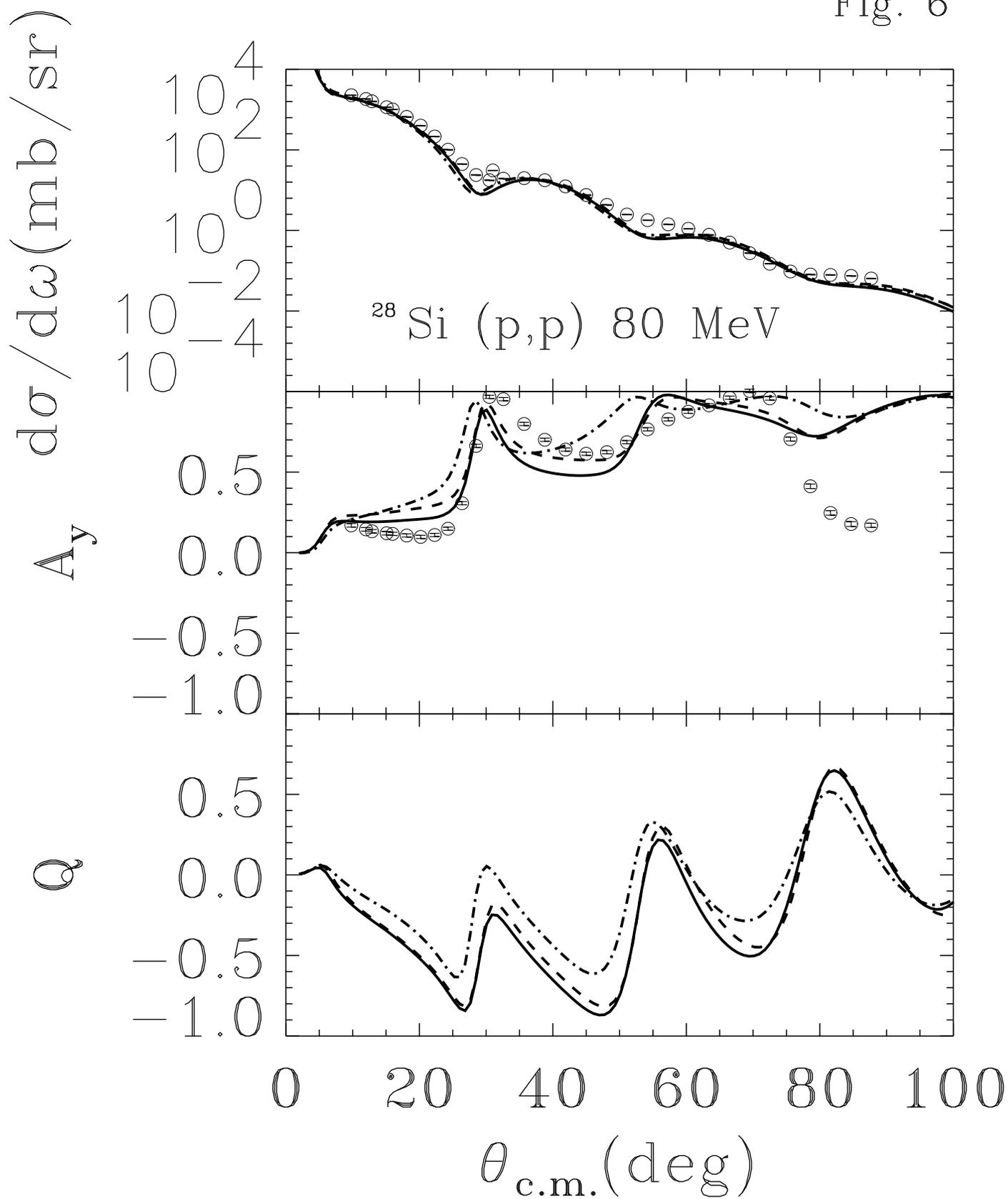
\begin{figure}
\caption{  Same as Fig.~2,  except for $^{28}$Si at 80~MeV proton
kinetic energy. The data are from Ref. \protect\cite{si80}. \label{fig6} }
\end{figure}

\noindent
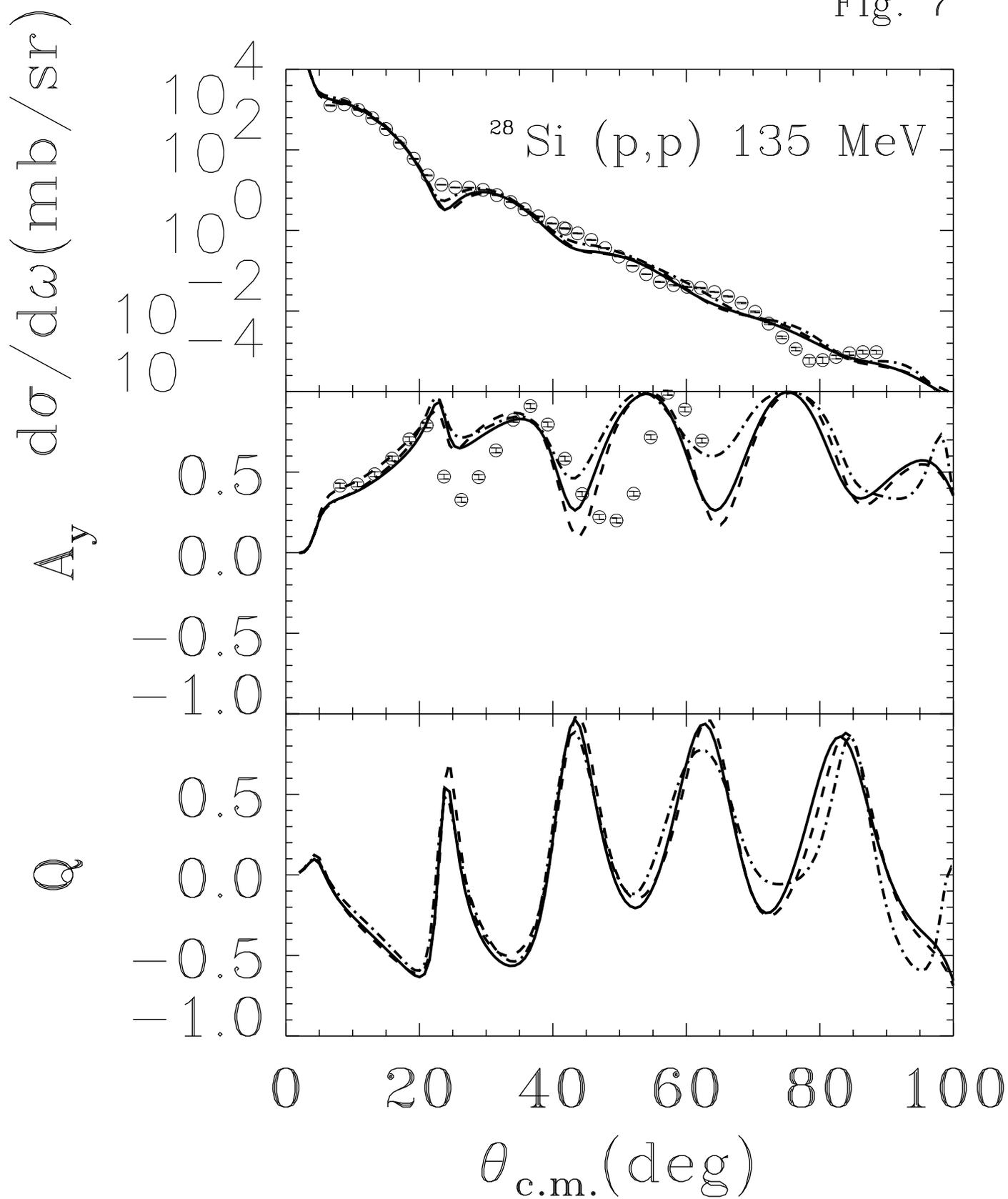
\begin{figure}
\caption{  Same as Fig.~6,  except the projectile kinetic energy is
135~MeV, and the data are from Ref. \protect\cite{si80,si135}.\label{fig7} }
\end{figure}

\noindent
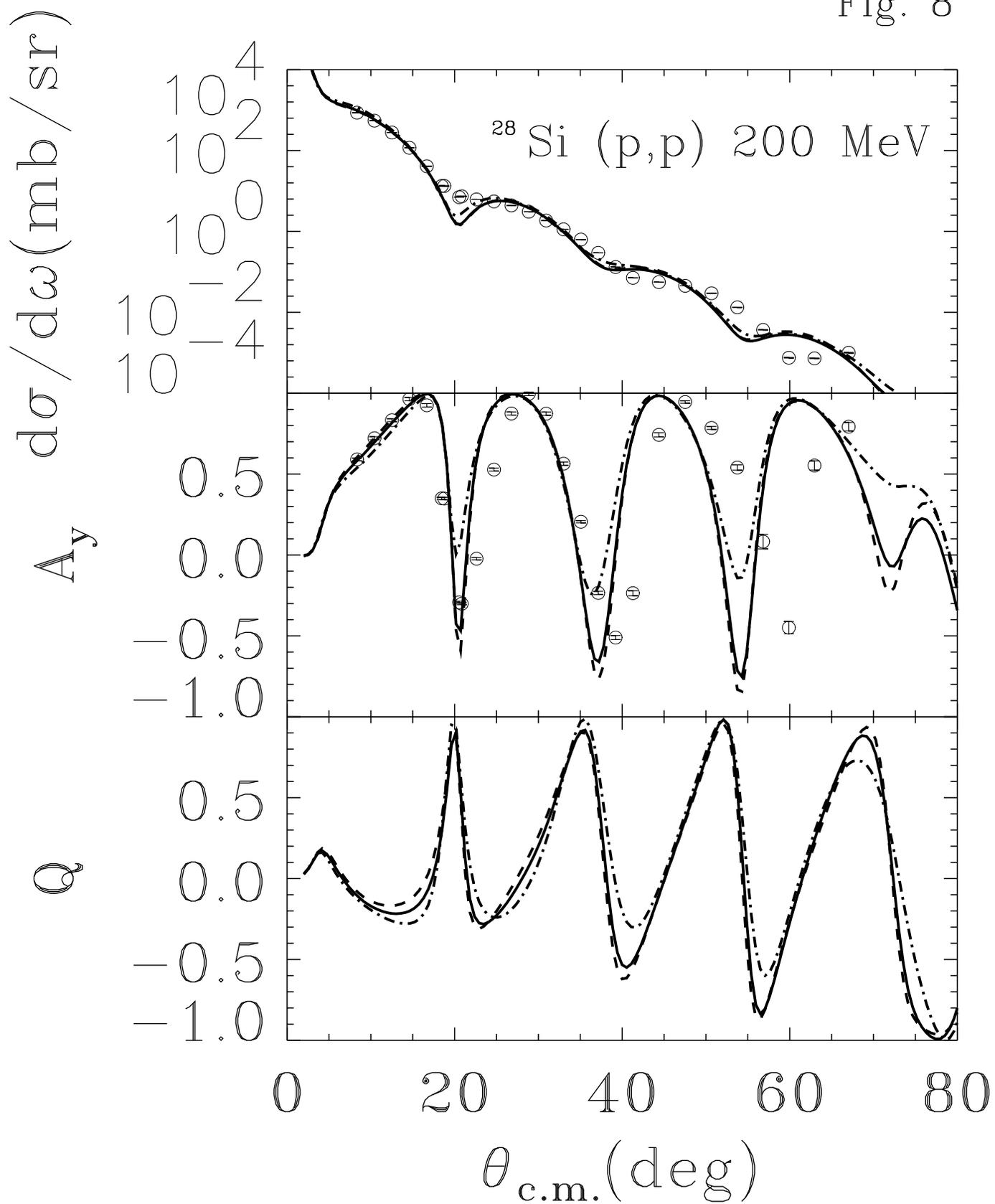
\begin{figure}
\caption{  Same as Fig.~6,  except the projectile kinetic energy is
200~MeV, and the data are from Ref. \protect\cite{si200}. \label{fig8} }
\end{figure}

\noindent
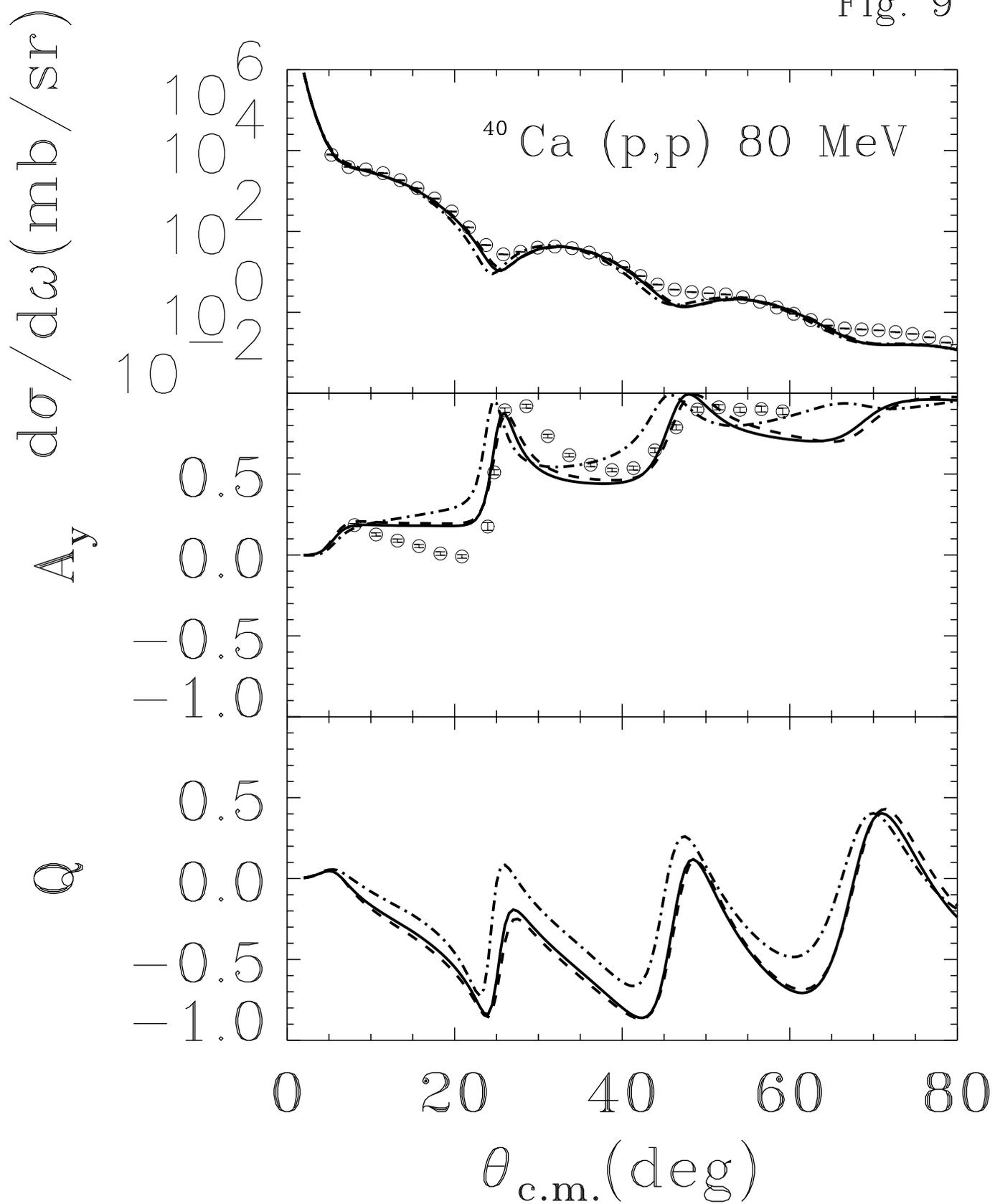
\begin{figure}
\caption{  Same as Fig.~2,  except for $^{40}$Ca at 80 MeV proton
kinetic energy. The data are from Ref. \protect\cite{ca80}. \label{fig9} }
\end{figure}

\noindent
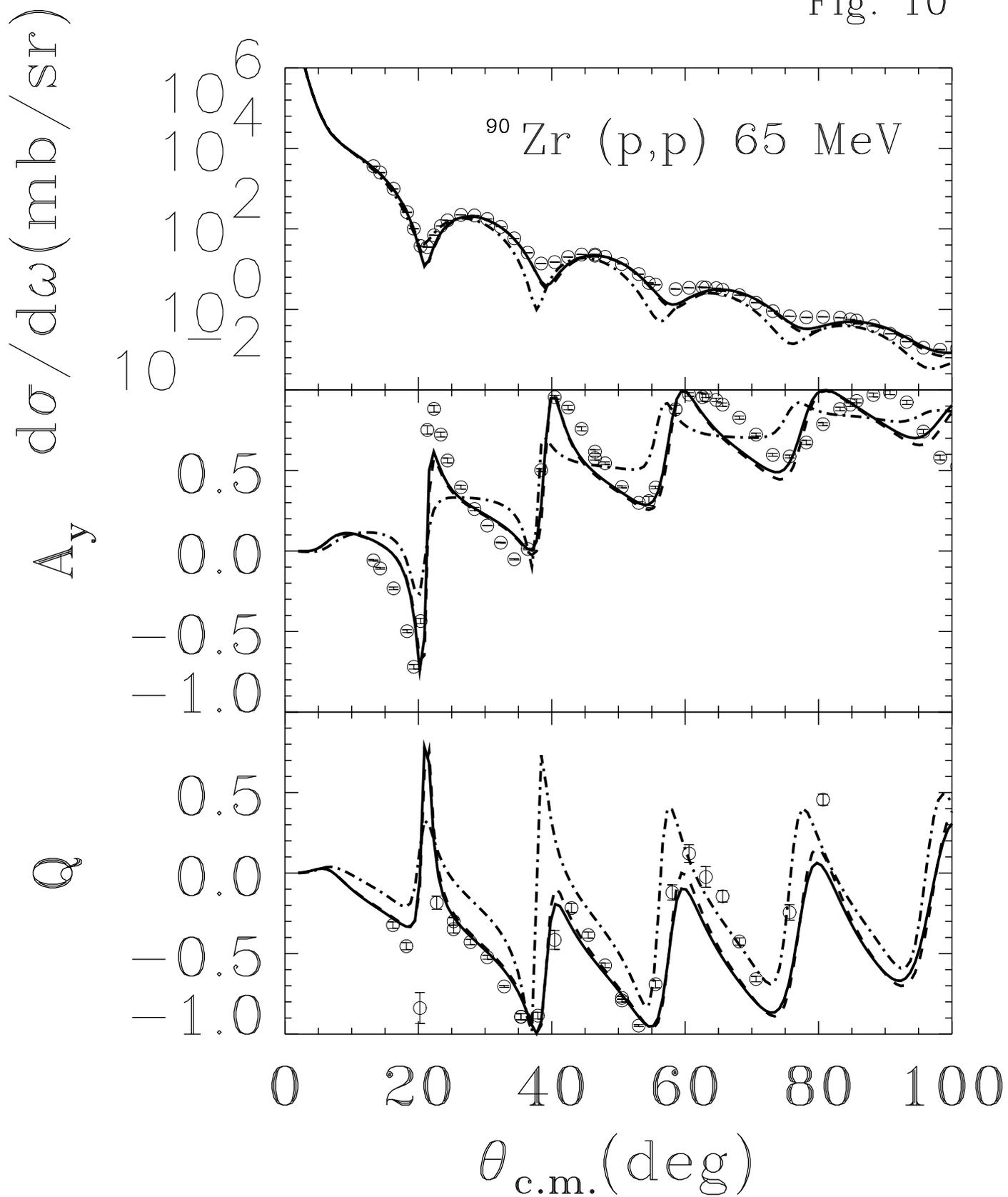
\begin{figure}
\caption{  Same as Fig.~2,  except for $^{90}$Zr at 65 MeV proton
kinetic energy. The data are from Ref. \protect\cite{zr65}.\label{fig10} }
\end{figure}

\noindent
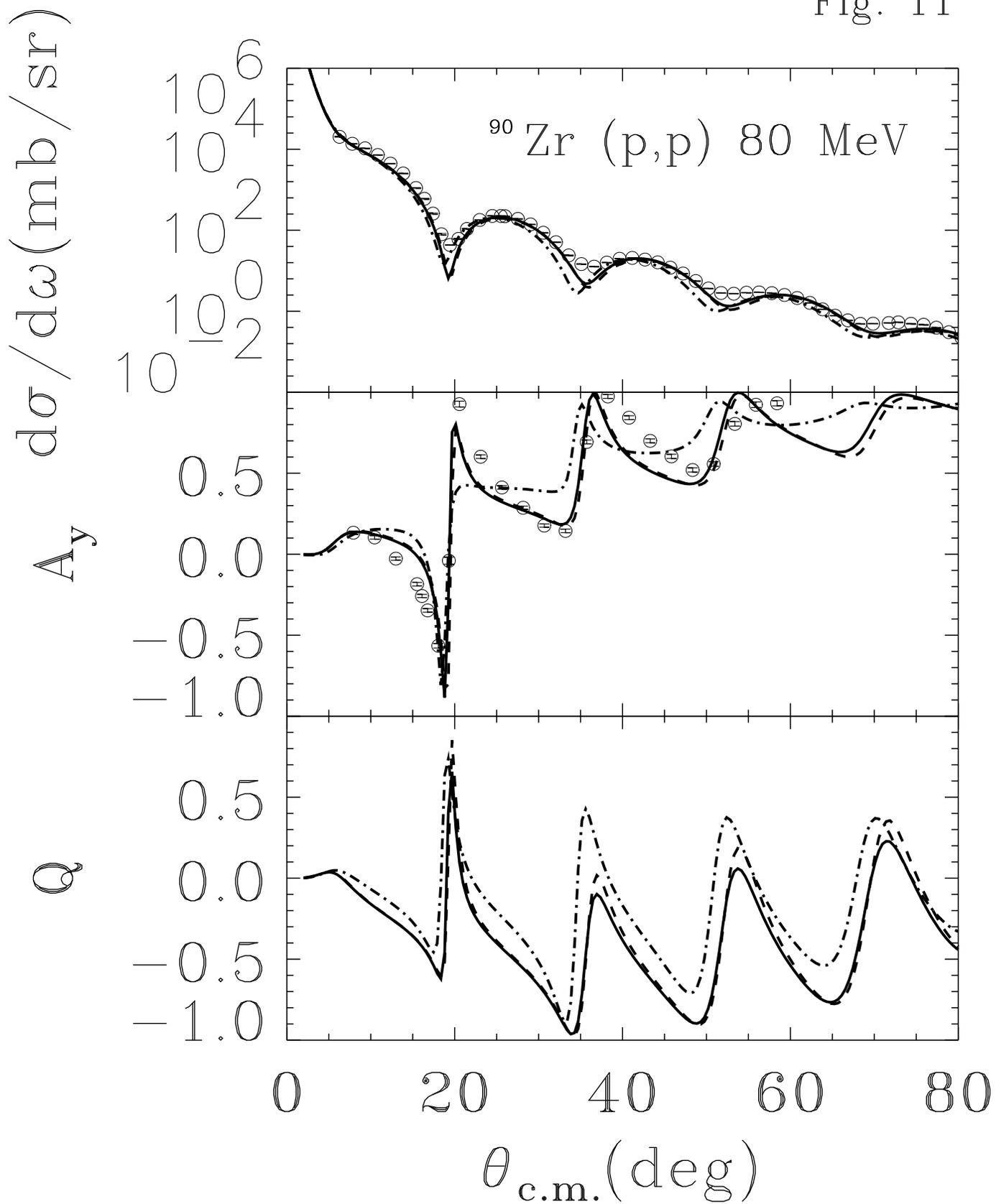
\begin{figure}
\caption{  Same as Fig.~10,  except for $^{90}$Zr at 80 MeV proton
kinetic energy. The data are from Ref.
\protect\cite{si135}.\label{fig11} }
\end{figure}

\noindent
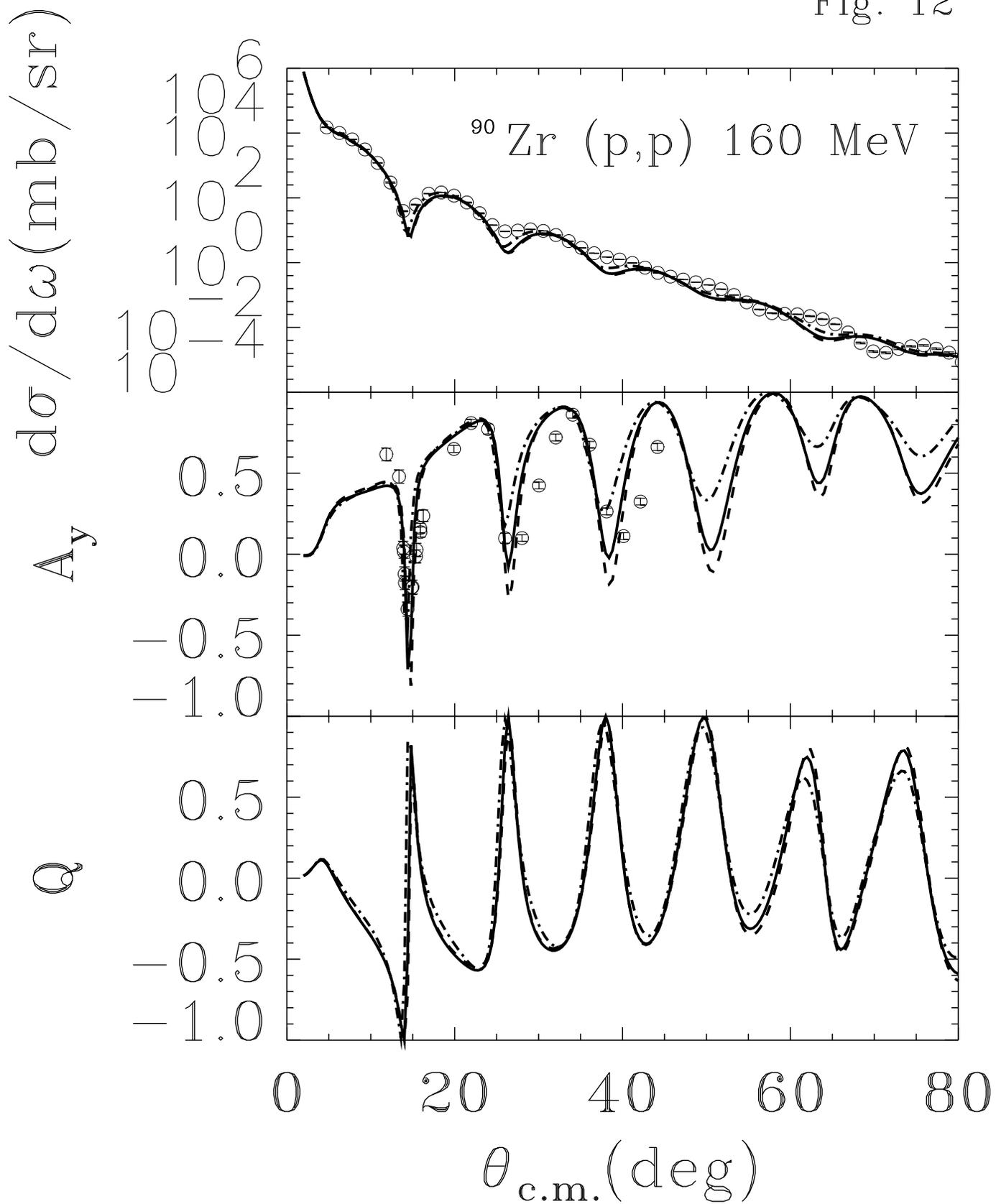
\begin{figure}
\caption{  Same as Fig.~10,  except except the projectile kinetic energy
is 160~MeV. The data are from Ref. \protect\cite{si135}. \label{fig12} }
\end{figure}

\noindent
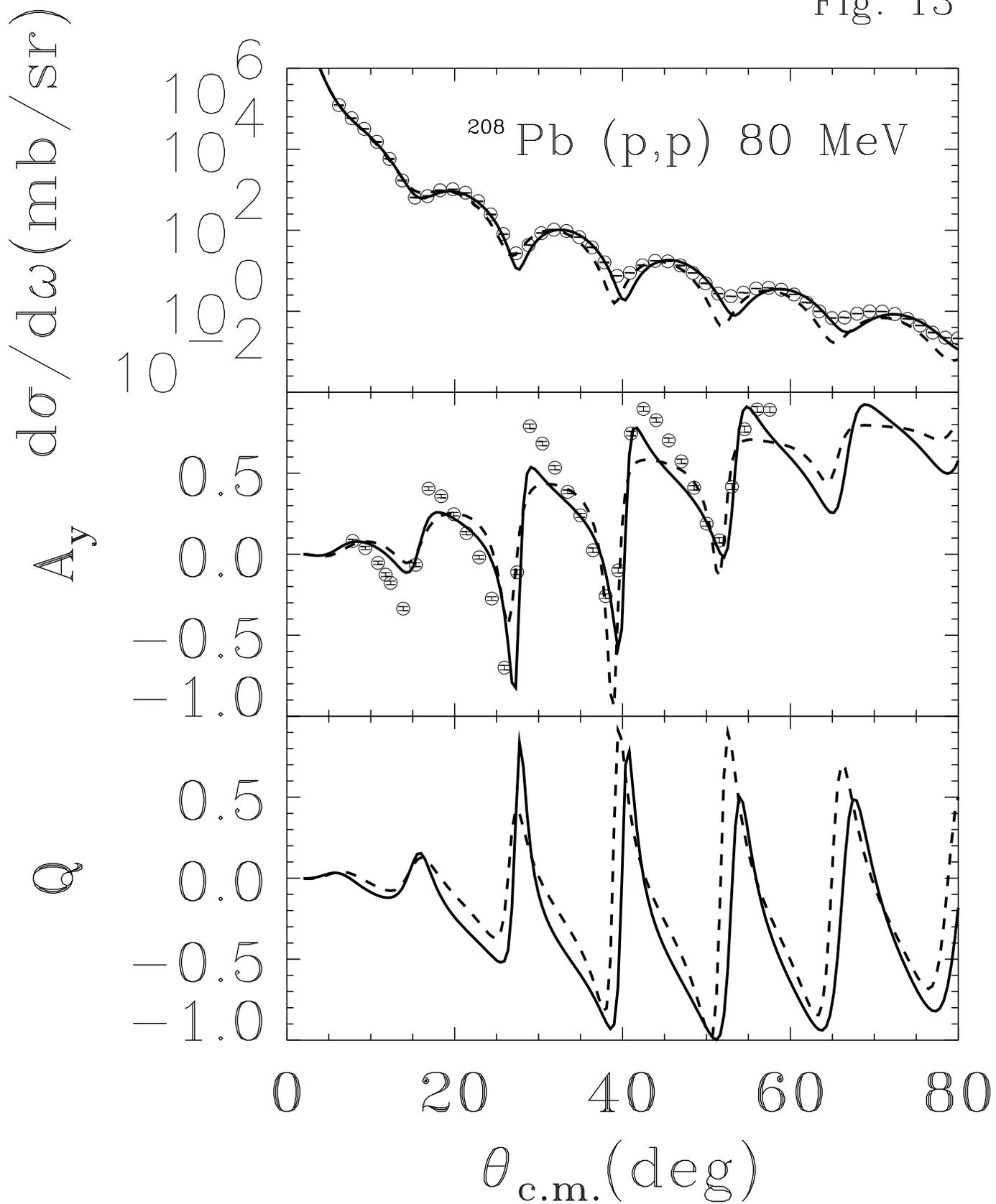
\begin{figure}
\caption{  Same as Fig.~2,  except for $^{208}$Pb at 80~MeV proton
kinetic energy. The data are from Ref. \protect\cite{ca80}. \label{fig13} }
\end{figure}

\noindent
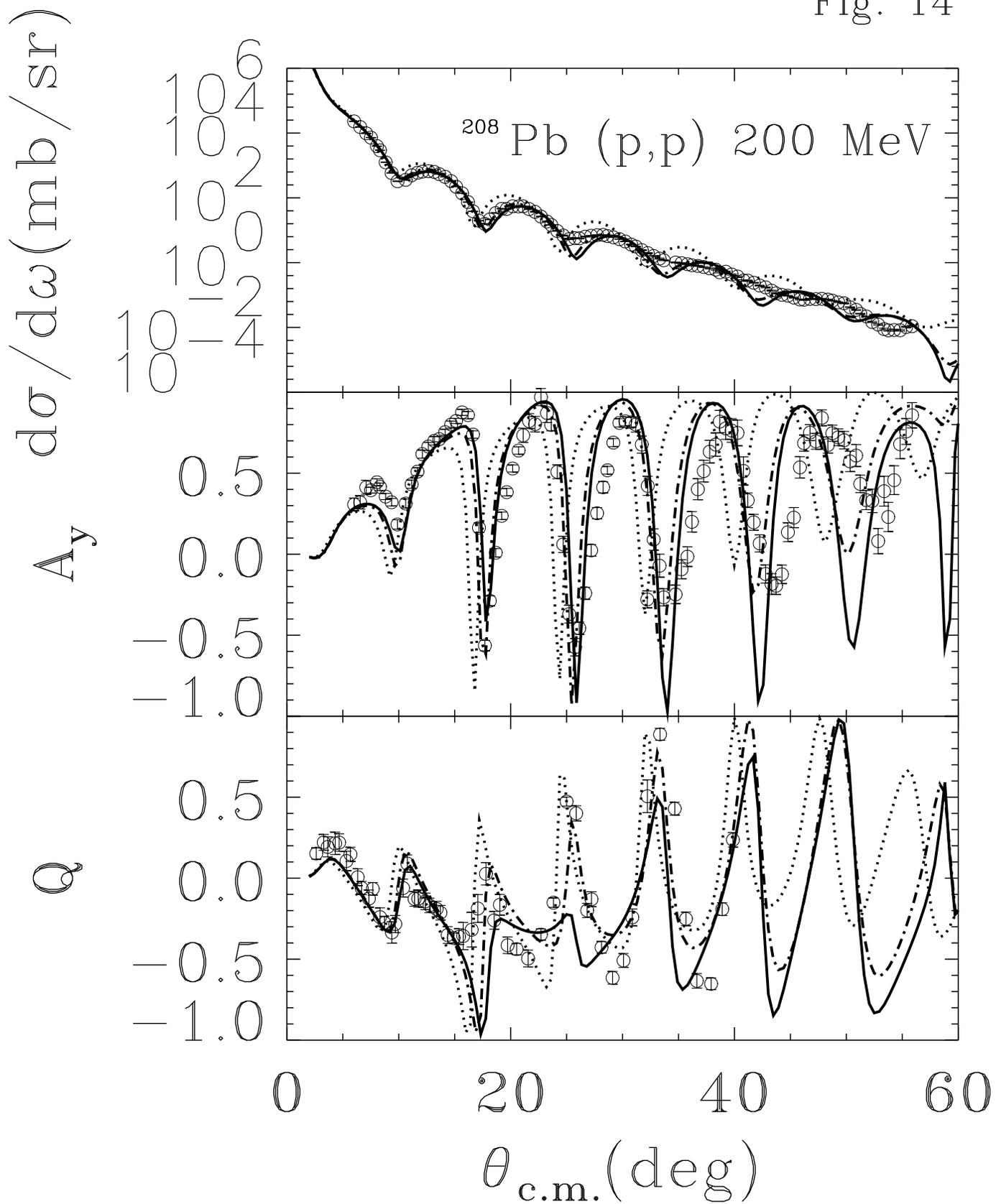
\begin{figure}
\caption{Same as Fig.~13,  except except the projectile kinetic energy
         is 200~MeV. The dotted line represents a local `on-shell
         $t\rho$' calculation which uses only the on-shell values of
         the same t-matrix. The data are from
         Ref.~\protect\cite{pb200}. \label{fig14}}
\end{figure}

\end{document}